\documentstyle[12pt]{article}
\begin{document}


\newcommand{\Z}{{\ \sf Z\hspace*{-1.7ex} \sf Z \ }}
\newcommand{\R}{{\sf R\hspace*{-1.5ex}\rule{0.18ex}{1.5ex}
      \hspace*{-0.2ex}\rule[1.42ex]{0.5ex}{0.15ex}
            \hspace*{-0.3ex}\rule[0.75ex]{0.5ex}{0.15ex}\hspace*{0.9ex}}}
\newcommand{\N}{{\sf N\hspace*{-1.65ex}\rule{0.18ex}{1.5ex}
      \hspace*{-0.2ex}\rule[1.42ex]{0.4ex}{0.15ex}\hspace*{1.0ex}}}
\newcommand{\Q}{{\sf Q\hspace*{-1.1ex}\rule{0.17ex}{1.5ex}\hspace*{1.1ex}}}
\newcommand{\C}{{\sf C\hspace*{-0.9ex}\rule{0.17ex}{1.4ex}\hspace*{0.9ex}}}


\begin{titlepage}

\baselineskip=20 pt

\begin{flushright}
   SU-HEP-4241-547\\
   March 1994\\
\end{flushright}

                             \vspace{2 cm}

                             \begin{center}
             \Large\bf Geometry of the Frenkel-Kac-Segal cocycle.\\

                            \vspace{2.5 cm}

\baselineskip=20pt

             \large\sc Nikolaos Kalogeropoulos $^{\dagger \ \star}$\\
                               \medskip
            \normalsize\sl Department of Physics\\ Syracuse University\\
                       Syracuse, NY 13244-1130\\
                              \end{center}

                             \vspace{2 cm}

                            \begin{center}
                          \large\sc ABSTRACT\\
                             \end{center}

\rm We present an analysis of the cocycle appearing in the vertex
operator representation of simply-laced, affine, Kac-Moody algebras.
We prove that it can be described in the context of  $R$-commutative
geometry, where $R$ is a Yang-Baxter operator, as a strong $R$-commutative
algebra. We comment on the Hochschild,
cyclic and dihedral homology theories that appear in non-commutative geometry
and their potential relation to string theory.

                                \vfill

\footnoterule
\noindent\footnotesize $^{\dagger}$ Research supported in part by the DOE grant
              DE-FG02-85ER40231\\
              $^{\star}$ e-mail \ \ nikolaos@suhep.phy.syr.edu

\end{titlepage}

                              \newpage


\baselineskip=20pt

                            \begin{center}
                     \large\sc 1.\ \ Introduction\\
                             \end{center}

\rm The vertex operator construction \cite{GORev}, \cite{GOMSRI}, \cite{Kac},
\cite{FLM} arose during the early days of string
theory. Although it was not known under this name and it was not made rigorous
until the mid-80's many of its essential ingredients were invented and
developed during the mid-70's \cite{Hal}. This construction provides a means
for
introducing interactions in the string world-sheet. It is known that
theories of fundamental strings
are free field theories on the world-sheet (something which is not true in
the case of effective string theories) in the sense that there is
no potential for the matter fields, just their kinetic terms as well as
coupling to the 2-dim metric (and other fields like the dilaton, the
antisymmetric tensor field etc). The need appeared at that time for a
mechanism that could provide interactions in the string world-sheet. This is
the vertex operator construction. From a spacetime viewpoint this machinery
allows us to describe emission and absorption of string states, bringing the
whole approach closer to that of the LSZ in the case of ordinary field
theories.\\

Apart from its physical motivation the vertex operator construction received a
lot of attention for its mathematical importance. The affine Kac-Moody algebras
\cite{Kac} are examples of infinite dimensional Lie algebras that are easy
enough to allow
an explicit study of their geometric significance as well as a study of their
representations. Geometrically they are central extensions of loop algebras.
{}From the viewpoint of representation theory they have many similar features
with the finite dimensional Lie algebras (as well as important differences).
The vertex operator calculus \cite{FLM} was developed as a tool for providing
representations for the affine Kac-Moody algebras. Although the analysis that
we will follow in the present paper is valid (in most of its points) for simply
laced algebras, the construction has been generalized to twisted Kac-Moody
algebras as well as to non simply-laced algebras \cite{GNOS} etc.\\

Non-commutative geometry \cite{Connes1} is the study of geometric objects that
are defined over non-commutative algebras. The commutative case is relatively
well understood, at many points, and it provides the mathematical framework for
algebraic quantum field theory \cite{Haag}. The non-commutative approach was
revived in the early 80's primarily through Connes' influential papers
\cite{Connes1}. At the same time (but independently)
cyclic homology theory was developed by Tsygan \cite{Tsygan},
Loday-Quillen \cite{LQ} and Connes \cite{Connes2}.
Although several applications of non-commutative geometry have appeared
in the physics literature we believe that its potential is far from having been
completely realized. One particular case of the non-commutative geometry is the
theory of quasi-triangular quasi-Hopf algebras \cite{D1}, \cite{D2} that has
received a lot of attention, primarily because of its importance in low
dimensional field theories \cite{Majid}.
In this paper we will discuss another part of non-commutative
geometry which like quantum groups has as one central object the study of some
kind of Yang-Baxter operators. In that sense it has been influenced by
low-dimensional field theories, although the constructions can be carried out
in any number of dimensions. Whether this type of mathematics will be
useful for better understanding physical systems is an open issue.\\

Let's summarize the content of the paper. In section 2 we give a brief outline
of the theory of vertex operators. We have followed closely the terminology
used in the physics literature instead of the one in representation theory.
In section 3 we examine some algebraic properties of the
Frenkel-Kac-Segal cocycle (FKS cocycle). We prove that a modified form of these
cocycles forms an associative, unital, involutive algebra over the commutative
ring $\Z$ of integers. In section 4 we discuss in general some homology
theories that help us analyze the structure of non-commutative algebras. We
point out some potential applications of these techniques for string theories.
In the same section we also prove that the FKS cocycles fit nicely in the
framework of $R$-commutative geometry. In section 5 we work out, partly, an
example that may help illustrate some of the points made in the previous
sections. In section 6 we present some conclusions of the current study as well
as some hints for potentially interesting directions of research. The Appendix
contains proofs of some properties that are essential in our treatment but
whose inclusion in the main text would impede the natural development of the
subject.\\

                                \vfill

                                \newpage


                             \begin{center}
          \large\sc 2.\ \ The Vertex Operator Construction\\
                              \end{center}

\rm The vertex operator construction provides a level one representation of
affine, untwisted, simply-laced Kac-Moody algebras. The construction has been
generalized to other cases as well, but we will not discuss them in this
paper. The exposition in this section follows closely that of \cite{GORev}.\\

A starting point of our investigation is the need to understand the
representation theory of loop groups. These groups appear very often in a
variety of physical systems, and share many characteristics of the finite
dimensional Lie groups which makes their analysis more tractable. One such
characteristic is the existence of an exponential mapping that provides a
connection between the affine, Kac-Moody algebra or a central extension of it
and its associated loop group or its central extension respectively. So, from
the viewpoint of representation theory it suffices to consider representations
of the affine, Kac-Moody algebras. We are interested in the heighest weight,
unitary, integrable representations only.\\

An affine Kac-Moody algebra which is simply-laced (the normalization is such
that the lenght of the roots is 2) and at level one, has generators $H^{i}_{m},
E^{\alpha}_{n}, \ m\in\Z ,\ i=1,\ldots ,n $ , $\alpha$ is an element of the
root system of {\bf g}, which in a Cartan-Weyl basis satisfy the relations
\begin{equation}
 [H^{i}_{m}, H^{j}_{n}] = m\delta^{ij}\delta_{m+n,0}
\end{equation}
\begin{equation}
 [H^{i}_{m}, E^{\alpha}_{n}] = \alpha^{i}E^{\alpha}_{m+n,0}
\end{equation}
\begin{equation}
 [E^{\alpha}_{m}, E^{\beta}_{n}] =
  \left \{
    \begin{array}{ll}
    \epsilon (\alpha, \beta ) E^{\alpha +\beta}_{m+n}
& if \ \alpha +\beta \ \ is \ a \ root \ of \ {\bf g}\\
    \alpha\cdot H_{m+n} +m\delta_{m+n,0} & if \ \ \alpha +\beta = 0\\
    0 & otherwise
    \end{array}
  \right \}
\end{equation}
where $\epsilon (\alpha, \beta ) = \pm 1$. We will discuss more extensively its
properties in the next section. We observe, to begin with, that
$ H^{i}_{m} , H^{i}_{-m} , \ i=1,\ldots , rank {\bf g}, \ n=1, 2 \ldots ,$
form a harmonic oscillator algebra. We may associate the superscript $i$ with
the spacetime degrees of freedom and the subscript $m$ with the different
oscillator modes of the string. In cases in which we have infinite quantities
of interest, it is occasionally helpful to consider the properties of their
generating function. In this case the generating function is the
Fubini-Veneziano momentum field, and using the usual conventions for labelling
the indices in the Taylor expansions, we may write
\begin{equation}
   H^{i}(z) = \sum_{n\in\Z} H^{i}_{-n}z^{n}
\end{equation}
where $z$ is a coordinate of the world-sheet which can be considered for our
purposes as an expansion parameter. Equation (4) can be written in the more
familiar form
\begin{equation}
P^{i}(z) = p^{i} +
  \sum_{n=1}^{\infty}(\alpha^{i}_{n}z^{-n} + \alpha^{i}_{-n}z^{n})
\end{equation}
where $p^i$ is the momentum of the center-of-mass of the string (which
coincides with the zeroth oscillatory mode) and the $\alpha$'s denote the
higher
modes. From (5) we find
\begin{equation}
 p^{i} = \oint_{C_0}\frac{dz}{2\pi i} P^{i}(z)z^{-1}
\end{equation}
where the integration is over the unit circle (or a homotopically equivalent
path) surrounding the origin once in the positive (counterclockwise) direction.
Then since $P^{i\dagger}(z) = P^{i}(z)$, (6) implies that
\begin{equation}
 p^{i\dagger} = p^{i}
\end{equation}
Similarly
\begin{equation}
 \alpha^{i}_{n} = \oint_{C_0}\frac{dz}{2\pi i} P^{i}(z)z^{n-1}
\end{equation}
and therefore
\begin{equation}
  (\alpha^{i}_{n})^{\dagger} = \alpha^{i}_{-n}
\end{equation}
So Hermitian conjugation connects the ``opposite" modes of oscillation of the
string that form the creation and annihilation operators in the quantized
string formalism.\\
The Fubini-Veneziano coordinate field $Q^{i}(z)$ is defined by the differential
relation
\begin{equation}
  P^{i}(z) = iz\frac{dQ^{i}}{dz}
\end{equation}
which gives
\begin{equation}
 Q^{i}(z) = q^{i}- ip^{i}lnz + \sum_{n\neq 0}\alpha^{i}_{n}\frac{z^{-n}}{n}
\end{equation}
where $q^{i}$ is an integration constant which is identified as the position of
the center of mass of the string  and it is canonically conjugate to $p^{i}$.
We
define the bosonic normal ordering $::$ by moving the $\alpha^{i}_{n}, \ n\geq
1$ to the right of $\alpha^{j}_{m}, \ m\leq -1$ and $p$ to the right of $q$.
Consider the operator
\begin{equation}
  U^{\alpha}(n) = z^{{\alpha}^{2}/2}:e^{i\alpha\cdot Q(z)}:
\end{equation}
where $\alpha$ denotes the vector whose components in the $rank \ {\bf g}$
dimensional spacetime are $\alpha^{i}$. We observe that equation (12) does not
have any singularities exactly because of the existence of the normal ordering.
The factor $z^{{\alpha}^{2}/2}$ has been introduced for convenience in
subsequent formulas. The vertex operator defined by equation (12) obeys the
equation
\begin{equation}
 [p^{i} , U^{\alpha}(z)] = \alpha^{i} U^{\alpha}(z)
\end{equation}
which means that the corresponding state $U^{\alpha}(z)|0>$ has momentum
$\alpha$ i.e. the operator $U^{\alpha}(z)$ introduces momentum $\alpha$ to the
string states on which it acts. In terms of its modes
\begin{equation}
  U^{\alpha}(z) = \sum_{m\in\Z} A^{\alpha}_{m}z^{-m}
\end{equation}
Although tedious, it is straightforward to prove that the modes
$A^{m}_{\alpha}$ obey the following relations with the Cartan subalgebra
generators , as well as among themselves
\begin{equation}
  [H^{i}_{m} , H^{j}_{n}] = m\delta^{ij}\delta_{m+n,0}
\end{equation}
\begin{equation}
  [H^{i}_{m} , A^{\alpha}_{n}] = \alpha^{i}A^{\alpha}_{m+n}
\end{equation}
\begin{equation}
A^{\alpha}_{m}A^{\beta}_{n}-(-1)^{\alpha\cdot\beta}A^{\beta}_{n}A^{\alpha}_{m}=
 \left\{
   \begin{array}{ll}
     A^{\alpha +\beta}_{m+n} & if \ \ \alpha\cdot\beta = -1\\
     \alpha\cdot H_{m+n} + m\delta_{m+n,0} & if \ \ \alpha\cdot\beta = -2\\
     0 & if \ \ \alpha\cdot\beta\geq 0
   \end{array}
 \right\}
\end{equation}
For a simply-laced algebra if $\alpha +\beta$ is a root then
$(\alpha +\beta)^2 = 2$ which, since $\alpha^2 = \beta^2 = 2$ gives
$\alpha\cdot\beta = -1$. Similarly if $\alpha +\beta =0$ then
$(\alpha +\beta )^2 = 0$ so $\alpha\cdot\beta = -2$. Finally, if
$\alpha\cdot\beta\geq 2$ then $(\alpha +\beta )^2 > 2$ i.e. $\alpha +\beta$ is
neither a root nor $\beta = -\alpha$. This establishes the one-to-one relation
between the right-hand sides of equations (3) and (17). By comparing the two
sets of equations (1), (2), (3) and (15), (16), (17) we see that the momentum
modes and the vertex operator modes almost provide a level-1 representation of
the affine, simply-laced, Kac-Moody algebra {\bf g}.
The two different points between these
two sets of equations are the existence on the left-hand side of equation (17)
of the factor $(-1)^{\alpha\cdot\beta}$ and the lack of the factor $\epsilon
(\alpha , \beta)$ in the first line of the same equation when compared to
equation (3). The origin of the factor $(-1)^{\alpha\cdot\beta}$ is due to the
fact that in proving (15)-(17) we have used the operator product expansion
\begin{equation}
  :e^{i\alpha\cdot Q(z)}::e^{i\beta\cdot Q(w)}: =
(z-w)^{\alpha\cdot\beta}:e^{i[\alpha\cdot Q(z) +\beta\cdot Q(w)]}: \
                                                           , \ \ \ |z|>|w|
\end{equation}
Then the term $(-1)^{\alpha\cdot\beta}$ is the scaling constant of
the singular term on the operator product expansion of two vertex operators.
Since we never allow two operators to occupy the same point, (18) is always
non-singular.\\

In order to exactly get a representation of the Kac-Moody algebra we introduce
the additional quantities $c_{\alpha}$  defined by \cite{FrKac}, \cite{Segal}
\begin{equation}
  E^{\alpha}_{m} = A^{\alpha}_{m}c_{\alpha} = c_{-\alpha}A^{\alpha}_{m}
\end{equation}
with the properties
\begin{equation}
  c^{\dagger}_{\alpha} = c_{\alpha} \ \ \ and  \ \ \ c^{2}_{\alpha}=1
\end{equation}
When this is the case we see that (15)-(17) form a representation of the
Kac-Moody algebra (1)-(3).\\

One more point to which we should pay attention is that the vertex operator
$U^{\alpha}(z)$ should be single-valued when $z$ encirccles the origin once.
The potentially troublesome term for the single-valuedness of $U^{\alpha}(z)$
is \ \ $z^{{\alpha}^2/2}e^{i\alpha\cdot q}z^{\alpha\cdot p}$.\ \
In order to have a single-valued expression the dependence of the vertex
operator on its arguments should be polynomial which means that
 \begin{equation}
      \alpha\cdot p + \frac{{\alpha}^{2}}{2} \in \Z
 \end{equation}
If the ground state has momentum $\bar{p}$ then upon acting with the vertex
operator $U^{\alpha}(z)$ we find another state of momentum
$p\in\Lambda_{R}+\bar{p}$ where $\Lambda_{R}$ is the root lattice of ${\bf g}$.
So any two possible values of $p$ differ by an element of the root lattice and
then (22) implies that
 $$\alpha\cdot\beta\in\Z \ \ \ if \ \ \beta\in\Lambda_{R}$$
which is the condition for $\Lambda_{R}$ to be an integral lattice. In the case
in which {\bf g} is simply-laced, $\Lambda_{R}$ is, in addition, even.\\


                             \begin{center}
         \large\sc 3.\ \ Algebraic properties of the fks cocycle\\
                              \end{center}

\rm In this section we present several identities that the Frenkel-Kac-Segal
cocycle satisfies. They form the basis for the cohomology theories that we
construct in the next two sections.\\

In order to avoid introducing any extra degrees of freedom in the construction
of the cocycles, we express them \cite{FrKac}, \cite{Segal},
\cite{GORev} as follows:  Let $\alpha$ , $\beta$ be elements
of the root lattice $\Lambda_{R}$. Suppose that the ground state has momentum
$\bar{p}$. Then we set
\begin{equation}
  c_{\alpha} = \sum_{\beta\in\Lambda_{R}} \epsilon (\alpha,\beta )
                                           |\beta + \bar{p}><\beta + \bar{p}|
\end{equation}
The states $|\beta + \bar{p}>$ are eigenstates of the momentum operator $p$
with
eigenvalues $\beta^{\mu}+\bar{p}^{\mu}$.
The functions $\epsilon (\alpha,\beta )$
take values in $\Z_{2}$. In order to have a representation of the Kac-Moody
algebra we require $\epsilon (\alpha,\beta )$ to obey the following relations
\begin{equation}
  \epsilon (\alpha,\beta ) = (-1)^{\alpha\cdot\beta + {\alpha}^2 {\beta}^2}
                               \epsilon (\beta,\alpha )
\end{equation}

\begin{equation}
  \epsilon (\alpha,\beta ) \epsilon (\alpha +\beta, \gamma ) =
      \epsilon (\alpha,\beta +\gamma ) \epsilon (\beta,\gamma )
\end{equation}

\noindent Here we want to point out two things. First, not all
lattices admit an
$\epsilon$ function satisfying these properties. However, the root lattices of
the simply-laced algebras (the ones belonging to the $A$, $D$, $E$ series in
the
Cartan classification) do admit one. 
Second, the choice of
$\epsilon (\alpha, \beta)$ is not unique. Indeed, suppose that
\begin{equation}
 \epsilon^{\prime} (\alpha,\beta ) =
            \eta_{\alpha}\eta_{\beta}\eta_{\alpha +\beta}\epsilon (\alpha,\beta
)
\end{equation}
with $\eta_{\alpha}= \pm 1$. It is easy to confirm that
$\epsilon^{\prime} (\alpha,\beta )$ also satisfies the equations (23), (24) if
$\epsilon (\alpha,\beta )$ does. It is probably worth thinking of this freedom
of redefinition of $\epsilon (\alpha,\beta )$ as a $\Z_2$ gauge invariance at
every point of the root lattice.  Subsequently we can use this freedom to ``fix
the gauge" in a manner that simplifies our calculations.\\

\noindent To begin with, putting $\beta = 0$ in (24) we see that
$$\epsilon (\alpha,0 ) \epsilon (\alpha,\gamma ) =
                             \epsilon (\alpha,\gamma) \epsilon (\beta,\gamma
)$$
Since all the $\epsilon$ functions commute (being numerical functions) this
implies that
\begin{equation}
              \epsilon (\alpha,0 ) = \epsilon (0,\gamma )
\end{equation}
By partially using the gauge freedom we can set $\eta_{0}=\epsilon (\alpha,0)$.
Then
\begin{equation}
        \epsilon^{\prime} (\alpha,\beta ) = \epsilon^{2} (\alpha,\beta ) = 1
\end{equation}
So, we have that
\begin{equation}
      \epsilon (\alpha,0) = \epsilon (0,\beta ) = 1
\end{equation}
This result has been obtained by partially using the gauge freedom that we had.
To fully fix the gauge we additionally require that
$\eta_{\alpha}\eta_{-\alpha} = \epsilon (\alpha, -\alpha)$ which gives
\begin{equation}
                  \epsilon (\alpha,-\alpha ) = 1
\end{equation}

\noindent We must make certain that these two conditions (28), (29) are
preserved in the
``future". In a constraint system (in the Dirac approach) this would mean that
the primary constraints should have vanishing Poisson bracket with the
Hamiltonian. In our case there is no Hamiltonian; we have presented this
analogy for additional clarity. Therefore the conditions (28), (29) cannot be
violated (or, in Dirac's language, there are be no secondary constraints).
Choosing $\alpha +\beta +\gamma = 0$ (24) gives
   $$\epsilon (\alpha,\beta )\epsilon (-\gamma,\gamma ) =
       \epsilon (\alpha,-\alpha )\epsilon (\beta,-\alpha -\beta )$$
which considering the gauge fixing condition (8) becomes
\begin{equation}
          \epsilon (\alpha,\beta ) = \epsilon (\beta,-\alpha -\beta )
\end{equation}
Repeating once more the above relation we find
\begin{equation}
    \epsilon (\beta,-\alpha -\beta ) = \epsilon (-\alpha -\beta, \alpha )
\end{equation}
Choosing $\alpha + \beta = 0$, (3) gives
$$\epsilon (\alpha,-\alpha )\epsilon (0,\gamma ) =
        \epsilon (\alpha,-\alpha +\gamma ) \epsilon (-\alpha,\gamma )$$
Due to the gauge fixing conditions and the fact that
$\epsilon^{2} (\alpha,\beta ) = 1$ we find
   $$\epsilon (\alpha,-\alpha +\gamma  ) = \epsilon (-\alpha,\gamma )$$
By setting $\beta = -\alpha +\gamma$ we get
\begin{equation}
  \epsilon (\alpha,\beta ) = \epsilon (-\alpha,\alpha +\beta )
\end{equation}
Combining (10) and (11) we see that
\begin{equation}
    \epsilon (\alpha,\beta ) = \epsilon (-\beta,-\alpha )
\end{equation}
In what follows we assume that the momentum eigenstates form a complete ,
orthonormal set in the Hilbert space of states, i.e.
\begin{equation}
            <\mu | \nu > =\delta_{\mu\nu}
\end{equation}
{}From now on the primary objects of interest are the modified cocycles
defined by
\begin{equation}
   \hat{c}_{\alpha} = e^{iq\alpha}c_{\alpha}
\end{equation}
   and they are thus given explicitly by
\begin{equation}
  \hat{c}_{\alpha} = e^{iq\alpha} \sum_{\beta\in\Lambda_{R}}
          \epsilon (\alpha,\beta ) |\beta + \bar{p}><\beta + \bar{p}| =
                      \sum_{\beta\in\Lambda_{R}} \epsilon (\alpha,\beta )
                             |\alpha +\beta +\bar{p}><\beta + \bar{p}|
\end{equation}
In the Appendix we prove that
\begin{equation}
  \hat{c}_{\alpha}\hat{c}_{\beta} =
                             \epsilon (\alpha,\beta ) \hat{c}_{\alpha + \beta}
\end{equation}
We define the multiplication operation in the space of cocycles to be the one
given by (16). Using the fact that
$\hat{c}_{\alpha + \beta} = \hat{c}_{\beta +\alpha}$ as well as (37) and (23)
we
find
\begin{equation}
\hat{c}_{\beta}\hat{c}_{\alpha} = (-1)^{\alpha\cdot\beta + {\alpha}^2
{\beta}^2}
                                   \hat{c}_{\alpha}\hat{c}_{\beta}
\end{equation}
Besides, the associativity property holds (see the Appendix for the proof)
\begin{equation}
  \hat{c}_{\alpha}(\hat{c}_{\beta}\hat{c}_{\gamma}) =
                        (\hat{c}_{\alpha}\hat{c}_{\beta})\hat{c}_{\gamma}
\end{equation}
Now consider the element
 $$\hat{c}_{0} = e^{iq0}\sum_{\beta\in\Lambda_{R}}
            \epsilon (0, \beta) |\beta + \bar{p}><\beta + \bar{p}|$$
We immediately see that due to our gauge fixing (7) as well as the completeness
of the momentum basis eigenfunction
\begin{equation}
             \hat{c}_{0} = 1
\end{equation}
Then we have from equation (37)
\begin{equation}
  \hat{c}_{0}\hat{c}_{\alpha} = \hat{c}_{\alpha}\hat{c}_{0} = \hat{c}_{\alpha}
\end{equation}
We also see that
\begin{equation}
 \hat{c}_{-\alpha} = e^{-iq\alpha}\sum_{\beta\in\Lambda_{R}}
                                           |\beta + \bar{p}><\beta + \bar{p}|
\end{equation}
satisfies the equation (with the gauge fixing condition (29))
\begin{equation}
  \hat{c}_{-\alpha}\hat{c}_{\alpha} = 1 = \hat{c}_{\alpha}\hat{c}_{-\alpha}
\end{equation}
Therefore we have seen that considering the objects $\hat{c}_{\alpha}$, where
$\alpha$ is an element of the root lattice, with the multiplication defined
by (37) we get a non-commutative group. We can extend, the previous
construction by introducing a second binary operation, the addition, between
the cocycles. It is obvious that the cocycles satisfy the associativity as well
as the commutativity property with respect to addition. When we add cocycles we
should pay attention to the fact that the ground states in their expansion in
momentum eigenstates (36) should have the same momenta. Otherwise we would be
adding elements belonging to different superselection sectors of the theory,
something which is not allowed. Apart from the addition we introduce the
multiplication of an integer and a cocycle. The definition is
\begin{equation}
  \lambda\hat{c}_{\alpha} = \lambda e^{iq\alpha}\sum_{\beta\in\Lambda_{R}}
              \epsilon (\alpha, \beta ) |\beta + \bar{p}><\beta + \bar{p}|
\end{equation}
where $\lambda\in \Z$. It is clear that the following distributive properties
hold for this multiplication
 $$(\lambda + \mu )\hat{c}_{\alpha} =
                             \lambda\hat{c}_{\alpha} + \mu\hat{c}_{\alpha}$$
 $$\lambda (\hat{c}_{\alpha} + \hat{c}_{\beta}) =
                    \lambda\hat{c}_{\alpha} + \lambda\hat{c}_{\beta}$$
for $\hat{c}_{\alpha}, \hat{c}_{\beta}$ cocycles and $\lambda, \mu$ integers.
We define the zero element of the addition (denoted by $\hat{o}$) to be
\begin{equation}
                     0\hat{c}_{\alpha} = \hat{o}
\end{equation}
We also have the opposite element with respect to addition defined by the
property
\begin{equation}
                 \hat{c}_{\alpha} + (-\hat{c}_{\alpha}) = \hat{o}
\end{equation}
The set of cocycles with the addition and the multiplication becomes a
non-commutative, associative, unital algebra. It is the group algebra of the
group of cocycles (having as group operation the multiplication). This group
algebra is defined over the ring of integers. We call the group $G$, and the
group algebra $\Z[G]$.\\

In addition to all the above properties, this algebra has an
involution. It is expressed by the condition
\begin{equation}
                 \hat{c}_{\alpha}^{\dagger} = \hat{c}_{-\alpha}
\end{equation}
The previous equation comes from the
requirement of constructing unitary representations of the Kac-Moody algebras.
Equation (47) as well as the fact that it is an involution of the group algebra
are proved in the Appendix.\\

\noindent As we noted at the beginning of the section, all the simply-laced
algebras admit a function $\epsilon$ with properties (23), (24). We have not
confined ourselves, to the treatment of even lattices only, since the
incorporation of fermions requires the inclusion in the formalism of odd
lattices as well.\\

In what follows we will consider only the algebraic properties of the
cocycles. In doing so, we follow the spirit of the approach of the algebraic
quantum field theorists \cite{Haag}. Their viewpoint is that all the essential
features of a system are incorporated in its algebraic structure. Therefore, we
must consider the algebra to be the starting point, and from that try to
determine its representations, the observables etc. of the quantum theory. Here
we adopt a similar approach.\\


                              \begin{center}
     \large\sc 4.\ \  Homology of and R-commutative geometry of \Z[G]\\
                               \end{center}

\rm We have noticed in the previous section that the algebra of the modified
FKS cocycles is group algebra. In this section we will comment on the homology
of this algebra and prove that it is an example of $R$-commutative
geometry \cite{Manin}, \cite{Baez1}.\\

Two homology theories that one frequently uses in  non-commutative geometry
are the Hochschild \cite{Hoch} and the cyclic homology theories \cite{Connes2},
\cite{LQ}, \cite{Tsygan}. In case that the algebra
has an involution (as in our case) dihedral homology is more pertinent in the
analysis \cite{Loday1}, \cite{Loday2}, \cite{Lodder}.
We discuss here some properties of these theories that allow us
to perform some computations in simple cases and suggest connections with
string theory.\\

Let $G$ be a group, $x\in G$ and $G_x$ be the centralizer i.e. the set
$G_{x} = \{ g\in G : gx=xg \}$ . The nerve of $G$, denoted by $B.G$ is defined
to be the simplicial set with objects $B_{n}G=G^{n}$  face maps
\begin{equation}
  d_{i}(g_{1},\ldots ,g_{n}) = \left\{
    \begin{array}{ll}
        (g_{2},\ldots ,g_{n}) & for \ \ \ i=0\\
        (g_{1},\ldots ,g_{i}g_{i+1},\ldots ,g_{n}) & for \ \ \ 1\leq i\leq
n-1\\
        (g_{1},\ldots ,g_{n-1} & for \ \ \ i=n-1
    \end{array}
  \right\}
\end{equation}
and degeneracy maps
\begin{equation}
s_{j}(g_{1},\ldots ,g_{n}) = (g_{1} ,\ldots ,g_{i},1,g_{i+1},\ldots ,g_{n})
\end{equation}
Here $(-1)^{n}$ is the sign of the cyclic permutation on $n+1$ letters.\\

Consider an element $x$ of the center of $G$ and define an action of the cyclic
operator $t_n$ on $B_{n}G$ by
\begin{equation}
 t_{n}(g_{1},\ldots ,g_{n}) = (x(g_{1}g_{2}\cdots g_{n})^{-1},
                              g_{1}, \ldots ,g_{n-1})
\end{equation}
The operation $t_n$ puts a cyclic structure on the nerve of $G$ that depends on
$x$ such that $t_{n}^{n+1}=1$. This cyclic space is denoted by $B.(G,x)$ and it
is called the twisted nerve of $(G,x)$. In the special case in which $x=1$ we
get a canonical cyclic structure on the nerve $B.G$. Denoting the Hochschild
homology by $HH_n$, the cyclic homology by $HC_n$ and the group homology by
$H_n$, one can prove that \cite{Bur}
\begin{equation}
 HH_{n}(\Z[G]) \simeq \oplus_{<x>\in <G>} H_{n}(\Z[B.(G_{x},x)])
               \simeq \oplus_{<x>\in <G>} H_{n}(G_{x})
\end{equation}
where $<x>$ denotes the conjugacy class of \ $x$ \ and $<G>$ denotes the set of
conjugacy classes of its elements. The homology functors on the right-hand side
of this equation denote group homology. In the special case in which $x=1$, \ \
$G_{x}=G$ so
\begin{equation}
  H_{n}(\Z[B.(G_{1}, 1)]) = H_{n}(G)
\end{equation}
Similar considerations apply for the cyclic homology. In that case we can prove
that \cite{Loday2}
\begin{equation}
 HC_{n}(\Z[G])\simeq \oplus_{<x>\in <G>} HC_{n}(\Z [B.(G_{x},x)])
\end{equation}
For the $<1>$ component we may also prove that
\begin{equation}
  HC_{n}(G) \simeq \oplus_{i\geq 0} H_{n-2i}(G)
\end{equation}
Equation (54) is a consequence of the horizontal periodicity of period 2 in the
bicomplex that defines cyclic homology. This is the reason why the indices of
the homology groups in the right-hand side of (54) differ by 2.\\

If $X$ is a cyclic set with $\Z[X]$ its associated group algebra, there is a
canonical isomorphism \cite{Jones}, \cite{Loday2}
\begin{equation}
     HC_{\ast}(\Z[X]) = H_{\ast}^{S^{1}}(|X|,\Z)
\end{equation}
where $|X|$ denotes the geometric realization of the cyclic set $X$ and the
functor $H_{\ast}^{S^{1}}$ is the $S^1$ equivariant homology of $|X|$ with
coefficients in $\Z$. In the case of group algebras we can do better than that
\cite{Bur}, \cite{Good}, \cite{Loday2}.
If $x$ is in the center of $G_x$, we define the Borel space $X(G_{x}, x)$ by
\begin{equation}
  X(G_{x},x) := ES^{1}\times_{S^1}BG_{x}
\end{equation}
where $ES^{1}$ is the universal bundle for $S^1$ and the $S^1$ action is
defined as follows: Let $\gamma_{x} :\Z\times G_{x}\rightarrow G_{x}$,
$\gamma_{x}(n,g)=x^{n}g$ be a group homomorphism. Then it induces the map
$B\gamma_{x}: S^{1}\times BG_{x}\rightarrow BG_{x}$ among ther classifying
spaces. We can prove that
\begin{equation}
  HC_{n}(\Z[G])\simeq \oplus_{<x>\in <G>} H_{n}(X(G_{x},x;\Z)
\end{equation}
Let's discuss the meaning of equation (57) from a more physical viewpoint. The
cyclic homology of a simplicial set, has already built in, through the action
of the cyclic operator $t_n$ an action which is similar to that of a circle.
Roughly speaking the $t_n$'s provide an action of the different points of a
simplicialization of the circle. Therefore for cyclic homology the fundamental
object which probes the structure of the theory is a circle. In equation (56)
we see that fact incorporated in the statement that the space $X(G_{x}, x)$ is
an equivariant $S^1$ space. This tempts us to believe that the cyclic homology
is an appropriate tool in examining the homology structure of string theory in
which reparametrization of the string by a constant angle would naturally lead
to considerations in $S^1$ equivariant homology. On the other hand (51), (52),
(53), (55) and (57) are strongly reminiscent of the Euler characteristic of an
orbifold. For that there are two definitions: the mathematicians \cite{HH}
define it as
\begin{equation}
  \chi (X/G) = \frac{1}{|G|}\sum_{g\in G} \chi (X^{g})
\end{equation}
where $X$ is a manifold on which a discrete group $G$, of order $|G|$, acts
with fixed points and $X^{g}$ is the fixed point set of $X$ under the action of
$g$. The string theorists \cite{DHVW} define the Euler characteristic as
\begin{equation}
  \chi (X, G) = \frac{1}{|G|}\sum_{gh=hg} \chi (X^{<g,h>})
\end{equation}
where $g$ and $h$ are elements of $G$ providing twisted boundary conditions for
the wavefunctions in the two directions of the string world-sheet and
$X^{<g,h>}$ is the fixed point set under the simultaneous action of $g$ and
$h$. From equation (59) when $g=1$ we recover equation (58). This corresponds
to considering the field theory limit (low energy limit) of string theory. The
role of
\begin{equation}
 HC_{n}(G) = HC_{n}(\Z[B.(G_{1}, 1)])
\end{equation}
can be seen in a similar manner. It expresses the ``untwisted" sector in cyclic
homology, which can be expressed in terms of the group homology (54). In string
theory though we have to consider, because of modular
invariance, all the twisted sectors. Similarly, cyclic homology
forces us to consider all the twisted nerves of the group by elements if its
center. The sum in (59) is over mutually commuting elements of $G$ and the sum
in (53) is over the conjugacy classes of elements of $G$, two sets whose
elements  are in one-to-one
correspondence. To conclude, we want to point out that cyclic homology (or a
numerical invariant, an ``Euler"-type characteristic defined from it) can be as
useful for describing string-associated quantities as singular homology (and
its associated Euler characteristic) is for describing field theories.\\

In this second part of this section we prove that the FKS cocycles provide an
example of strong $r$-commutative algebras \cite{Manin}, \cite{Baez1}.\\

In our analysis, so far, we have used the fact that the algebra of the FKS
cocycles is an associative, unital, involutive algebra over the integers. There
is one more property that has not being used yet, and this is
   \begin{equation}
          \hat{c}_{\alpha}\hat{c}_{\beta} = (-1)^{\alpha\cdot\beta +
            {\alpha}^{2}{\beta}^{2}}\hat{c}_{\beta}\hat{c}_{\alpha}
   \end{equation}
This expression strongly reminds us a supercommutativity property. A big
difference is that our algebra $\Z[G]$ does not have any natural grading,
therefore it is impossible to use the formalism of supergeometry to probe the
structure of interest. Evidently a generalization of the supergeometry is
required and $R$-commutative geometry provides such a framework. Here the
central object is a Yang-Baxter operator $R$ which provides a generalized
twist map that replaces the commutation or anticommutation of two elements of
the algebra.\\

Let $V$ denote the vector space in which the root lattice $\Lambda_{R}$ can
be considered as naturally embedded, i.e.
 $$V = \{ \sum_{i=1}^{r} m_{i}\alpha_{i} ,
                \ \ m_{i}\in\R , \alpha_{i}\in\Lambda_{R} \}$$
Consider an element $R\in End(V\otimes V)$, and define
 \begin{equation}
    R(\hat{c}_{\alpha}\otimes\hat{c}_{\beta}) =
(-1)^{\alpha\cdot\beta +{\alpha}^{2}{\beta}^{2}}
                                       \hat{c}_{\beta}\otimes\hat{c}_{\alpha}
 \end{equation}
$R$ is invertible, the inverse being
 $$R^{-1}(\hat{c}_{\beta}\otimes\hat{c}_{\alpha}) =
                  (-1)^{\alpha\cdot\beta +{\alpha}^{2}{\beta}^{2}}
                                       \hat{c}_{\alpha}\otimes\hat{c}_{\beta}$$
Besides, it is straightforward to check that $R$ satisfies the Yang-Baxter
equation
\begin{equation}
                  R_{12}R_{23}R_{12} = R_{23}R_{12}R_{23}
\end{equation}
on $V\otimes V\otimes V$. Indeed
$$R_{12}R_{23}R_{12}
   (\hat{c}_{\alpha}\otimes\hat{c}_{\beta}\otimes\hat{c}_{\gamma}) =
     (-1)^{\alpha\cdot\beta +\alpha\cdot\gamma +\beta\cdot\gamma
 + {\alpha}^{2}{\beta}^{2} + {\alpha}^{2}{\gamma}^{2} +
{\beta}^{2}{\gamma}^{2}}
     \hat{c}_{\gamma}\otimes\hat{c}_{\beta}\otimes\hat{c}_{\alpha}$$
and the same is true for $R_{23}R_{12}R_{23}$. Since
\begin{equation}
 R^{2}(\hat{c}_{\alpha}\otimes\hat{c}_{\beta}) =
                     \hat{c}_{\alpha}\otimes\hat{c}_{\beta}
\end{equation}
i.e. $R^2 = 1$ since $\Lambda_{R}$ is an integral lattice. In addition
\begin{equation}
  R(\hat{c}_{0}\otimes\hat{c}_{\alpha}) = \hat{c}_{\alpha}\otimes\hat{c}_{0}
\ \ \ and \ \ \
R(\hat{c}_{\alpha}\otimes\hat{c}_{0}) = \hat{c}_{0}\otimes\hat{c}_{\alpha}
\end{equation}
is true.
The multiplication property of the $\hat{c}_{\alpha}$'s that we have already
considered is a map $ m: \Z[G]\times\Z[G]\rightarrow\Z[G] $ given by
\begin{equation}
      m(\hat{c}_{\alpha} , \hat{c}_{\beta}) = \epsilon (\alpha, \beta )
                          \hat{c}_{\alpha +\beta}
\end{equation}
To formulate the next property we define the element $s_{nm}\in B_{n+m}$ given
by
$$s_{nm} = (s_{m}\cdots s_{1})(s_{m+1}\cdots s_{2})\cdots
                                                   (s_{n+m+1}\cdots s_{n})$$
where \ \ $s_{i} , \ \ 1\leq i\leq n$. \ \
Each factor (string of braid group generators inside each parenthesis) has $m$
elements and there are $n$ such factors. Pictorially this equation represents
the exchange of the $n$ rightmost strands of a braid with $n+m$ strands, with
the $m$ leftmost ones. Let $\rho : B_{n}\rightarrow V^{\otimes n}$ be a
representation of the braid group $B_{n}$ on $V^{\otimes n}$. Then we can
verify (see Appendix) that
\begin{equation}
  (m\otimes m)\rho (s_{22}) = R(m\otimes m)
\end{equation}
where both sides of this equation represent maps
 $ (\Z[G])^{\otimes 4}\rightarrow (\Z[G])^{\otimes 2} $. A Yang-Baxter operator
$R$ on the algebra $\Z[G]$ that satisfies (63), (65), (67) is called an
$R$-structure
on $\Z[G]$. Moreover, equation (64) is satisfied, so $R$ is a strong
$r$-structure on $\Z[G]$ and then $\Z[G]$ is a strong $r$-algebra. In addition
\begin{equation}
  m(\hat{c}_{\alpha} , \hat{c}_{\beta})
          = \epsilon (\alpha, \beta ) \hat{c}_{\alpha +\beta}
          = mR(\hat{c}_{\alpha}\otimes\hat{c}_{\beta})
\end{equation}
so $\Z[G]$ is a strong, $R$-commutative algebra.\\


                             \begin{center}
                      \large\sc 5.\ \ An Example\\
                              \end{center}

\rm In this section we provide an example which we work on some detail, in
order to illustrate several points that we made in the previous section.\\

We are working with the Lie algebra $su(3)$, which is denoted by $A_2$ in the
Cartan classification. It is a simple, simply-laced algebra of rank 2 and
dimension 8. This means that the number of roots in its root system is 6. They
are
\begin{center}
   \begin{tabular}{ccc}
$\alpha_{1} = ( \sqrt{\frac{1}{2}} , \sqrt{\frac{3}{2}} )$ &
  $\alpha_{2} = ( \sqrt{\frac{1}{2}} , -\sqrt{\frac{3}{2}} )$ &
       $\alpha_{3} = ( \sqrt{2} , 0 )$ \\
& & \\
$\alpha_{4} = ( -\sqrt{\frac{1}{2}} , -\sqrt{\frac{3}{2}} )$ &
  $\alpha_{5} = ( -\sqrt{\frac{1}{2}} , \sqrt{\frac{3}{2}} )$ &
       $\alpha_{6} = ( -\sqrt{2} , 0 )$\\
   \end{tabular}
\end{center}

\noindent
Out of these six roots $\alpha_{1}$ and $\alpha_{2}$ are the simple roots. The
other roots can be expressed in terms of these two in the following way
$\ \ \alpha_{3} = \alpha_{1} + \alpha_{2} , \ \ \alpha_{4} = -\alpha_{1} ,
\ \ \alpha_{5} = -\alpha_{2} , \ \ \alpha_{6} = -\alpha_{3}$.\\
The group of interest is
$$G = \{\pm \hat{c}_{\alpha}, \ \ \alpha\in\Lambda_{R} \}$$
with the binary operation being the multiplication of $\hat{c}_{\alpha}$'s.\\
The Cartan matrix is
\begin{equation}
 K_{ij} = \left(
     \begin{array}{cc}
       2 & -1\\
       -1 & 2
     \end{array}
            \right)
\end{equation}
It is symmetric and all the diagonal elements are equal to $2$ something
typical of the fact that $su(3)$ is simply laced.\\

\noindent To find the conjugacy classes suppose that $N$ is a normal subgroup
and that
$\hat{c}_{\beta}\in N, \ \hat{c}_{\alpha}\in G$. Then
\ \ $\hat{c}_{\alpha}\hat{c}_{\beta}\hat{c}_{-\alpha}\in N$.\ \
However
$$\hat{c}_{\alpha}\hat{c}_{\beta}\hat{c}_{-\alpha} = \epsilon (\alpha ,\beta )
 \epsilon (\alpha +\beta , -\alpha ) \hat{c}_{\beta}$$
and
$$\epsilon (\alpha +\beta , -\alpha ) = (-1)^{-\alpha\cdot (\alpha +\beta )}
                                          \epsilon (-\alpha, \alpha +\beta )$$
as well as
$$\epsilon (-\alpha, \alpha +\beta ) = \epsilon (\alpha , \beta )$$
Therefore
$$\hat{c}_{\alpha}\hat{c}_{\beta}\hat{c}_{-\alpha} =
                                       (-1)^{\alpha\cdot\beta}\hat{c}_{\beta}$$
we conclude that the conjugacy class of $\hat{c}_{\beta}$ is
$\pm\hat{c}_{\beta}$ depending on whether $\alpha\cdot\beta$ is even or odd
respectively. Then, the group $G$ has the decomposition
      $$G = \cdots\otimes \Z_{2}\otimes\cdots\otimes \Z_{2}\otimes\cdots $$
where the elements in the direct product are in one-to-one correspondence with
the elements of the root lattice $\Lambda_{R}$.
This also implies that the set of conjugacy
classes is isomorphic to the root lattice.\\

To compute the Hochschild homology of $G$ we use the formula (51)
 $$HH_{n} (\Z[G]) = \oplus_{<x>\in <G>} H_{n}(G_{x})$$
where $x\in\Lambda_{R}$ and $G_{x}$ is the normal subgroup of $x$. Let $x =
\hat{c}_{\alpha}$. The normal subgroup of $\hat{c}_{\alpha}$ is found by
requiring $\hat{c}_{\beta}\hat{c}_{\alpha}\hat{c}_{-\beta} = \hat{c}_{\alpha}$.
This amounts to requiring $(-1)^{\alpha\cdot\beta} = 1$ for all $\beta\in
G_{\hat{c}_{\alpha}}$ where $G_{\hat{c}_{\alpha}}$ is the normal subgroup of
$\hat{c}_{\alpha}$. Let $\beta = m_{1}\alpha_{1} + m_{2}\alpha_{2}$ with
$m_{1} , m_{2} \in \Z$ where $\alpha_{1}, \alpha_{2}$ are the simple roots of
$\Lambda_{R}$ and $\alpha = n_{1}\alpha_{1} + n_{2}\alpha_{2}$ with $n_{1} ,
n_{2} \in \Z$ fixed. Then $(-1)^{n_{1}m_{1}+n_{2}m_{2}} = 1$ \ \ i.e.  \ \
$n_{1}m_{1}+n_{2}m_{2}$ \ \ is even. This means that
               $$m_{2} =\frac{2k-n_{1}m_{1}}{n_2}$$
with $k\in \Z$ should be an integer. We do not know how to find the most
general
solution to
this equation. Let's concentrate though in one specific case that illustrates
the argument. Setting $\alpha = \alpha_{1}$ and then $\alpha = \alpha_{2}$ we
find for the normal subgroups
$$G_{\hat{c}_{\alpha_{1}}} = \{ \pm\hat{c}_{\gamma}\in G : \gamma =
                               m\alpha_{1} + 2k\alpha_{2}, \ \ m,k\in \Z \}$$
$$G_{\hat{c}_{\alpha_{2}}} = \{ \pm\hat{c}_{\delta}\in G : \delta =
                               2l\alpha_{1} + n\alpha_{2}, \ \ l,n\in \Z \}$$
This statement means that each one of the $G_{\hat{c}_{\alpha_{1}}}$ and
$G_{\hat{c}_{\alpha_{2}}}$ is a lattice
$$\Lambda_{\hat{c}_{\alpha_{1}}} = \{ m\alpha_{1} + 2k\alpha_{2},\ \ m,k\in \Z
\}$$
and
$$\Lambda_{\hat{c}_{\alpha_{2}}} = \{ 2l\alpha_{1} + n\alpha_{2},\ \l,n\in \Z
\}$$
having one $\Z_{2}$ group at each lattice site. So
$$G_{\hat{c}_{\alpha_{1}}} = \cdots\otimes \Z_{2}\otimes\cdots\otimes
\Z_{2}\otimes\cdots $$ with the number of factors being in one-to-one
correspondence with the elements of the lattice
$\Lambda_{\hat{c}_{\alpha_{1}}}$ and similarly for $G_{\hat{c}_{\alpha_{2}}}$.
The problem of computing the Hochschild homology of the group algebra reduces
then to computing the group homology of $G_{\hat{c}_{\alpha_{1}}}$, \
$G_{\hat{c}_{\alpha_{2}}}$ etc for every lattice point $\alpha_{i}$.
To achieve that, we use iteratively the K\"{u}nneth factorization theorem,
which
adapted to our case states that (in the case of two factors in the direct
product)
$$ H_{n}(\Z_2\otimes \Z_2) = H_{p}(\Z_2)\otimes H_{n-p}(\Z_2)\oplus
                                           Tor(H_{n}(\Z_2) , H_{n-p-1}(\Z_2))$$
Using a resolution for $\Z_2$ we find for its group homology
 $$ H_{p}(\Z_2) = \left\{
       \begin{array}{lll}
         \Z & if & p=0 \\
         \Z_{2} & if & p:odd \\
         0 & if & p:even , p>0 \\
       \end{array}\right\} $$
{}From basic homological algebra we also know that
$$Tor(\Z,\Z) = Tor(\Z_{m},\Z) = Tor(\Z,\Z_{m}) = 0 \ \ \ \ \ \
                                 Tor(\Z_{m},\Z_{n}) = \Z_{(m,n)}$$
where $(m,n)$ denotes the
greatest common divisor of the positive integers $m$ and $n$. Applying these
results, we find that the first two homology groups are
\begin{eqnarray*}
  H_{0}(\Z_2\otimes \Z_2) & = & \Z\otimes \Z\\
  H_{1}(\Z_2\otimes \Z_2) & = & \Z\otimes \Z\oplus \Z_{2}\otimes \Z
\end{eqnarray*}
We proceed in the same way in calculating $H_{2}(\Z_2\otimes \Z_2)$ etc and
applying iteratively these formulas in the K\"{u}nneth factorization we find
$H_{n}(\cdots\otimes \Z_2\otimes\cdots\otimes \Z_2\otimes\cdots )$.
The final result is the group homology of $G_{\hat{c}_{\alpha_{1}}}$.
Similarly we find the group
homology of all the normal subgroups corresponding to the elements of $G$.
By adding these results we find the Hochschild homology of the group algebra
$\Z[G]$.\\

Now we come to the computation of the cyclic homology of $\Z[G]$. The
fundamental ingredient of our analysis is, once more, the set of normal
subgroups $G_{\hat{c}_{\alpha_{i}}}$ corresponding to the different elements of
the root lattice $\Lambda_{R}$. The Borel space associated to
$G_{\hat{c}_{\alpha_{i}}}$ is
$$X(G_{\hat{c}_{\alpha_{i}}}, \hat{c}_{\alpha_{i}}) = ES^{1} \times_{S^{1}}
                                              BG_{\hat{c}_{\alpha_{i}}}$$
Since
$$G_{\hat{c}_{\alpha_{i}}} =
                   \cdots\otimes \Z_{2}\otimes\cdots\otimes \Z_2\otimes\cdots$$
then
$$BG_{\hat{c}_{\alpha_{i}}} =
           \cdots\otimes B\Z_{2}\otimes\cdots\otimes B\Z_2\otimes\cdots$$
given that $B\Z_2 = \R P^{\infty}$ we find
$$X(G_{\hat{c}_{\alpha_{i}}}, \hat{c}_{\alpha_{i}}) = ES^{1} \times_{S^{1}}
(\cdots\otimes \R P^{\infty}\otimes\cdots\otimes \R P^{\infty}\otimes\cdots )$$
Therefore we have to compute the $S^1$ equivariant homology groups
$$H_{n}^{S^1} (\cdots\otimes \R P^{\infty}\otimes\cdots\otimes
                                    \R P^{\infty}\otimes\cdots ; \Z )$$
which we will not attempt here.
A simpler cyclic homology group to compute is the cyclic homology of the group
$G$ itself. $G$ is the conjugacy class of the element $\hat{c_0}$ (i.e. the
unit) and it can be straightforwardly proved that there is a canonical
isomorphism (54)
               $$HC_{n}(G) \cong \oplus_{i\geq 0} H_{n-2i}(G)$$
as we have pointed out before
where the right hand side denotes the group homology of $G$. We have shown in
the first part of this section how to calculate $H_{p}(G),\ \ 1\leq p\leq n$.
The only difference here is that due to the nature of the cyclic homology we
have to consider only half of the homology groups. As we remarked in the
previous section this result may be interesting in getting better insight
about the structure of string theory.\\

We have previously remarked it is the dihedral homology which is better suited
in
dealing with involutive algebras. Unfortunately, we are not able to present any
explicit results for our example since we are working over the integers in
which 2 is not invertible and no known formula for such a case is either
proved or conjectured \cite{Loday1} ,\cite{Lodder}.\\

Regarding the Yang-Baxter operator $R$ that we are using in the formalism of
$R$-commutative geometry we have to say that in our case it is given by
       $$R(\hat{c}_{\alpha}\otimes\hat{c}_{\beta}) = (-1)^{K_{\alpha\beta}}
                                       \hat{c}_{\beta}\otimes\hat{c}_{\alpha}$$
where $K_{\alpha\beta}$ is the Cartan matrix given by (69) in the case
in which $\alpha$, $\beta$ are simple roots or a linear combination of its
elements otherwise.\\

                             \begin{center}
             \large\sc 6.\ \ Discussion and Conclusions\\
                             \end{center}

\rm In the first part of this paper we have shown how we may construct a level
1 representation of a simply-laced, affine, Kac-Moody algebra. We begin from
its Cartan subalgebra which generates a rank{\bf g} dimensional maximal torus.
The
moments of the vertex operator $U^{\alpha}(z)$ play the role of raising and
lowering operators for the weights of the states in that representation. A
discrepancy between the commutation relations of the Cartan-Weyl basis of the
algebra {\bf g} and those of the corresponding representation forces us to
consider a $\Z_{2}$ extension of the root lattice. This realizes the dependence
of the singular part of the operator product expansion of two vertex operators
on the difference of their arguments. A manifestation of the existence of this
$\Z_{2}$ extension is the existence of the cocycle $\hat{c}_{\alpha}$. It is
exactly the vertex operator representation of an affine Kac-Moody algebra that
allows us to establish the equivalence of the bosonic and fermionic
descriptions of the WZW models. By using the Coulomb gas formalism it has
been proved that any 2 dimensional WZW model can be reformulated as a theory
of free bosons coupled to background charges at infinity. The vertex operators
play a fundamental role in this construction.\\

In the second part of this paper we discussed how the (modified) FKS cocycles
can be described in the framework of $R$-commutative geometry. We have
presented,
to some extent, an analysis of the Hochschild as well as of the cyclic homology
of the group algebra $\Z[G]$. We have seen that due to the involution
(Hermitian conjugation) that this algebra comes equipped with, it is dihedral
homology that depicts most of the algebraic characteristics of that structure.
Since the ground ring $\Z$ does not contain the inverse element of 2, we have
been unable to specify the exact part that the dihedral homology plays as a
part of the cyclic homology. Presumably, using a spectral sequence-type
argument for the action of the dihedral group on the algebra $\Z[G]$ of the
cyclic bar complex we should be able to compute the dihedral homology even when
2 is a non-invertible element in the $\Z$ ring. Alternatively, we may want to
work with quaternionic homology \cite{Loday1}
since it is naturally equipped with a length 4
horizontal periodicity on its defining bicomplex, but even in that case there
is no known canonical isomorphism between the dihedral and the quaternionic
homology groups if 2 is not invertible in the ground ring. \\

It would be, probably, instructive to compare the cyclic homology of the FKS
cocycles with the Lie algebra homology of the corresponding Lie algebras. In
view of the already known relation between these two theories \cite{LP} in the
case of algebras of matrices, it could help us illustrate, even more, the
connection between a Lie group and the set of its representations.\\

Finally, we have commented, in passing, on the potential role of the cyclic
homology for string theory (see also \cite{GJP}). We have suggested that,
since the cyclic homology has a simplicialization of a circle already built in,
it may be more pertinent in describing several string characteristics than the
singular theory. It could
also be worth developing the theory of braided Hochschild homology that was
introduced by Baez in \cite{Baez2}. We have seen lately a lot of activity in
the
categorial description of physical models and we believe that the braided
Hochschild homology may provide an additional tool in further refining the
formalism which, among other things, can also be applied, in future, to
physical systems.\\


                             \vspace{0.5 cm}

\noindent\underline{Acknowledgement} \ \ \
I am very grateful to John Baez for sending me a copy of his unpublished work
on ``Hochschild Homology in a Braided Tensor Category" and for his comments on
the manuscript.\\


                             \begin{center}
                        \large\sc\ \ Appendix\\
                             \end{center}

\rm  In this Appendix we collect proofs of some equations,
that have been omitted from the main text but may be helpful in making the
paper more transparent.\\

\noindent\sl{A.\ \ Proof of equation (37)}\\ \rm
\noindent\rm We have
\begin{eqnarray*}
 \hat{c}_{\alpha}\hat{c}_{\beta} & = &
e^{iq\alpha}\sum_{\gamma\in\Lambda_{R}} \epsilon (\alpha, \gamma )
     |\gamma +\bar{p}><\gamma +\bar{p}|
e^{iq\beta}\sum_{\delta\in\Lambda_{R}} \epsilon (\beta, \delta )
     |\delta +\bar{p}><\delta +\bar{p}|\\
 & = & \sum_{\gamma\in\Lambda_{R}} \epsilon (\alpha, \gamma )
                     |\alpha +\gamma +\bar{p}><\gamma +\bar{p}|
            \sum_{\delta\in\Lambda_{R}} \epsilon (\beta, \delta )
                          |\beta +\delta +\bar{p}><\delta +\bar{p}|\\
 & = & \sum_{\gamma\in\Lambda_{R}}\sum_{\delta\in\Lambda_{R}}
           \epsilon (\alpha, \gamma ) \epsilon (\beta, \delta )
              |\alpha +\gamma +\bar{p}><\delta +\bar{p}|
                 \delta_{\gamma, \beta +\delta}\\
 & = & \sum_{\gamma\in\Lambda_{R}} \epsilon (\alpha, \gamma)
              \epsilon (\beta,\gamma -\beta ) |\alpha +\gamma +\bar{p}>
                                                    <\gamma -\beta +\bar{p}|\\
\end{eqnarray*}
\noindent Using (3), the last equation becomes
$$\hat{c}_{\alpha}\hat{c}_{\beta} = \sum_{\gamma\in\Lambda_{R}}
    \epsilon (\alpha, \beta )\epsilon (\alpha +\beta, \gamma -\beta )
      |\alpha +\gamma +\bar{p}><\gamma -\beta +\bar{p}|$$
Setting $\gamma - \beta = \delta$
$$\hat{c}_{\alpha}\hat{c}_{\beta} = \sum_{\gamma\in\Lambda_{R}}
          \epsilon (\alpha, \beta )\epsilon (\alpha +\beta , \delta )
                          |\alpha +\beta +\delta +\bar{p}><\delta +\bar{p}|$$
This can be written as
$$\hat{c}_{\alpha}\hat{c}_{\beta} = \sum_{\gamma\in\Lambda_{R}}
     \epsilon (\alpha, \beta )\epsilon (\alpha +\beta , \delta )
        e^{iq(\alpha +\beta)} |\delta +\bar{p}><\delta +\bar{p}|$$
which gives equation (37)
$$\hat{c}_{\alpha}\hat{c}_{\beta} =
                            \epsilon (\alpha, \beta ) \hat{c}_{\alpha +\beta}$$

\noindent\sl{B.\ \ Proof of equation (39)}\\
\noindent\rm By repeatedly using equation (37) we find
 $$\hat{c}_{\alpha}(\hat{c}_{\beta}\hat{c}_{\gamma}) =
      \hat{c}_{\alpha}\epsilon (\beta, \gamma )\hat{c}_{\beta +\gamma} =
          \epsilon (\beta, \gamma )\epsilon (\alpha, \beta +\gamma )
             \hat{c}_{\alpha +\beta +\gamma}$$
On the other hand by using (24) we have
 $$\hat{c}_{\alpha}(\hat{c}_{\beta}\hat{c}_{\gamma}) =
     \epsilon (\alpha, \beta )\epsilon (\alpha +\beta, \gamma )
        \hat{c}_{\alpha +\beta +\gamma} =
           (\hat{c}_{\alpha}\hat{c}_{\beta})\hat{c}_{\gamma}$$

\noindent\sl{C.\ \ Proof of equation (47)}\\ \rm
\begin{eqnarray*}
   \hat{c}_{\alpha}^{\dagger} & = &
           {(e^{iq\alpha}\sum_{\beta\in\Lambda_{R}}\epsilon (\alpha ,\beta )
                |\beta +\bar{p}><\beta +\bar{p}|)}^{\dagger}\\
  & = & \sum_{\beta\in\Lambda_{R}}\epsilon (\alpha ,\beta )
                |\beta +\bar{p}><\alpha +\beta +\bar{p}|
\end{eqnarray*}
Setting $\alpha +\beta = \gamma$ we find
$$\hat{c}_{\alpha}^{\dagger} = e^{-iq\alpha}\sum_{\gamma\in\Lambda_{R}}
   \epsilon (\alpha, \gamma -\alpha ) |\gamma +\bar{p}><\gamma +\bar{p}|$$
and using equation (24),
$$\hat{c}_{\alpha}^{\dagger} = e^{-iq\alpha}\sum_{\gamma\in\Lambda_{R}}
               \epsilon (-\alpha, \gamma )|\gamma +\bar{p}><\gamma +\bar{p}| =
                               \hat{c}_{-\alpha}$$

\noindent\sl{D.\ \ Proof of equation (67)} \rm
\begin{eqnarray*}
     R(m\otimes m) (\hat{c}_{\alpha}\otimes\hat{c}_{\beta}
                 \otimes\hat{c}_{\gamma}\otimes\hat{c}_{\delta})
    & = & \epsilon (\alpha , \beta )\epsilon (\gamma ,\delta )
          R(\hat{c}_{\alpha +\beta}\otimes\hat{c}_{\gamma +\delta})\\
    & = & \epsilon (\alpha , \beta )\epsilon (\gamma ,\delta )
         (-1)^{(\alpha +\beta )\cdot (\gamma +\delta ) + {(\alpha +\beta )}^{2}
         {(\gamma +\delta )}^{2}}\hat{c}_{\gamma +\delta}\otimes
            \hat{c}_{\alpha +\beta}
\end{eqnarray*}
On the other hand since $\rho$ is a representation of the braid group $B_4$
$$\rho (s_{22}) = \rho [(s_{2}s_{1})(s_{3}s_{2})] = R_{23}R_{12}R_{34}R_{23}$$
Then
$$\rho (s_{22})
    (\hat{c}_{\alpha}\otimes\hat{c}_{\beta}
      \otimes\hat{c}_{\gamma}\otimes\hat{c}_{\delta}) =
        (-1)^{(\alpha +\gamma )\cdot (\beta +\delta )
           + {\alpha}^{2}{\gamma}^{2} + {\alpha}^{2}{\delta}^{2} +
             {\beta}^{2}{\gamma}^{2} + {\beta}^{2}{\delta}^{2}}
                \hat{c}_{\gamma}\otimes\hat{c}_{\delta}
                   \otimes\hat{c}_{\alpha}\otimes\hat{c}_{\beta}$$
Given that $\ \alpha , \ \beta , \ \gamma , \ \delta \ $ are vectors of an
integral lattice we see tnat
$$ (-1)^{2\alpha\cdot\gamma + 2\alpha\cdot\delta + 2\beta\cdot\gamma +
   2\beta\cdot\delta} = 1$$
which allows us to write
$$\rho (s_{22})
    (\hat{c}_{\alpha}\otimes\hat{c}_{\beta}\otimes
       \hat{c}_{\gamma}\otimes\hat{c}_{\delta}) =
         (-1)^{(\alpha +\beta )\cdot (\gamma +\delta ) + {(\alpha +\beta )}^{2}
          {(\gamma +\delta
)}^{2}}\hat{c}_{\gamma}\otimes\hat{c}_{\delta}\otimes
             \hat{c}_{\alpha}\otimes\hat{c}_{\beta}$$
and consequently
$$(m\otimes m)\rho (s_{22})
(\hat{c}_{\alpha}\otimes\hat{c}_{\beta}\otimes
\hat{c}_{\gamma}\otimes\hat{c}_{\delta}) =
(-1)^{(\alpha +\beta )\cdot (\gamma +\delta ) + {(\alpha +\beta )}^{2}
  {(\gamma +\delta )}^{2}}\epsilon (\alpha, \beta )\epsilon (\gamma , \delta )
  \hat{c}_{\gamma +\delta}\otimes\hat{c}_{\alpha +\beta}$$
This proves equation (67).\\



\end{document}